\documentclass[10pt,twocolumn,letterpaper]{article}

\usepackage{iccv}
\usepackage{times}
\usepackage{epsfig}
\usepackage{graphicx}
\usepackage{amsmath}
\usepackage{amssymb}

\usepackage{authblk}

\usepackage[breaklinks=true,bookmarks=false]{hyperref}

\iccvfinalcopy 

\def\iccvPaperID{****} 

\ificcvfinal\fi

\begin{document}

\title{CAMEL: A Weakly Supervised Learning Framework for Histopathology Image Segmentation}

\author{Gang Xu$^1$, Zhigang Song$^2$, Zhuo Sun$^3$, Calvin Ku$^3$, Zhe Yang$^1$, Cancheng Liu$^3$, Shuhao Wang$^{1,3}$, Jianpeng Ma$^{4*}$, Wei Xu$^{1*}$\\
$^1$Tsinghua University $^2$The Chinese PLA General Hospital $^3$Thorough Images $^4$Fudan University \\
{\tt\small {xug14@mails.tsinghua.edu.cn, songzhg301@139.com, \{zhuo.sun,calvin.j.ku,liucancheng,eric.wang\}@thorough.ai}, jpma@fudan.edu.cn, \{yangzhe2017,weixu\}@tsinghua.edu.cn}
}

\maketitle
\ificcvfinal\thispagestyle{empty}\fi

\renewcommand{\thefootnote}{\fnsymbol{footnote}}

\footnotetext[1]{Corresponding author.}

\begin{abstract}


Histopathology image analysis plays a critical role in cancer diagnosis and treatment. To automatically segment the cancerous regions, fully supervised segmentation algorithms require labor-intensive and time-consuming labeling at the pixel level. In this research, we propose CAMEL, a weakly supervised learning framework for histopathology image segmentation using only image-level labels. Using multiple instance learning (MIL)-based label enrichment, CAMEL splits the image into latticed instances and automatically generates instance-level labels. After label enrichment, the instance-level labels are further assigned to the corresponding pixels, producing the approximate pixel-level labels and making fully supervised training of segmentation models possible. CAMEL achieves comparable performance with the fully supervised approaches in both instance-level classification and pixel-level segmentation on CAMELYON16 and a colorectal adenoma dataset. Moreover, the generality of the automatic labeling methodology may benefit future weakly supervised learning studies for histopathology image analysis.

\end{abstract}

\section{Introduction}

Histopathology image analysis is the gold standard for cancer detection and diagnosis. In recent years, the development of deep neural network has achieved many breakthroughs in automatic histopathology image classification and segmentation~\cite{Yi2018Cancer,Liu2017Detecting,Madabhushi2016Image}. These methods highly depend on the availability of a large number of pixel-level labels, which are labor-intensive and time-consuming to obtain.

To relieve the demand for theses fine-grained labels, people have proposed many weakly supervised learning algorithms only requiring coarse-grained labels at the image-level~\cite{Jia2017Constrained,Xu2014Deep,Xu2014Weakly}. However, due to the lack of sufficient supervision information, the accuracy is much lower than their fully supervised counterparts. One way to improve the performance of weakly supervised learning algorithms is to add more supervision constraints. For natural images, some studies~\cite{Dai2015BoxSup,Anna2016Simple,Lin2016ScribbleSup} have proven the effectiveness of adding bounding boxes or scribble information artificially in their weakly supervised learning process. CDWS-MIL~\cite{Jia2017Constrained} has also shown the advantage of artificial area constraints for weakly supervised histopathological segmentation. However, it still takes much effort to obtain artificial constraints, especially in histopathology, where only well-trained pathologists can distinguish the cancerous regions from the normal ones. Therefore, automatically enriching labeling information instead of introducing artificial constraints before building the segmentation model is crucial for weakly supervised learning.

\begin{figure*}[t]
\begin{center}
\includegraphics[width=0.8\textwidth]{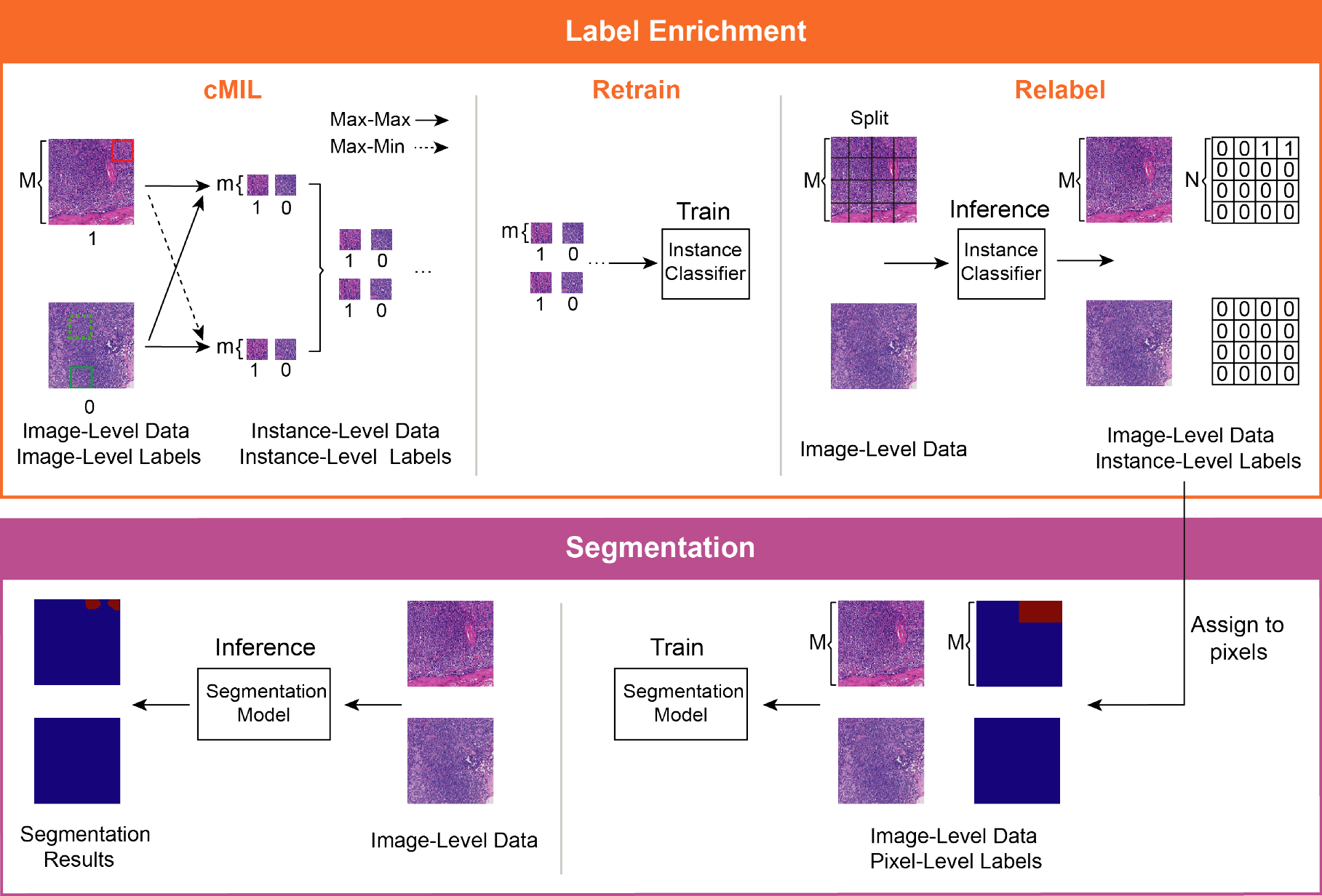}
\caption{System architecture of CAMEL. CAMEL consists of two basic steps: label enrichment and segmentation. M and m represent the size of the image and the instance, respectively. $N$ is the \emph{scale factor} of cMIL where $N=\frac{M}{m}$.}
\label{f1}
\end{center}
\end{figure*}


In this paper, we propose a weakly supervised learning framework, CAMEL, for histopathology image segmentation using only image-level labels. CAMEL consists of two steps: label enrichment and segmentation (Fig. \ref{f1}). Instead of introducing more supervision constraints, CAMEL splits the image into latticed \emph{instances} and automatically generates their instance-level labels in the label enrichment step, which can be regarded as a solution for a weakly supervised classification problem. In the label enrichment step, we use a combined multiple instance learning (cMIL) approach to construct a high-quality instance-level dataset with instance-level labels from the original image-level dataset. Then, we train a fully supervised classification model using this instance-level dataset. Once the model is trained, we split the images in the original image-level dataset into latticed instances and use this model to generate their labels. After label enrichment, the instance-level labels are directly assigned to their corresponding pixels, producing the approximate pixel-level labels and making fully supervised training of segmentation models possible. We conducted our experiments on CAMELYON16~\cite{Camelyon2016,Ehteshami2017Diagnostic} and a colorectal adenoma dataset, the results of both instance-level classification and pixel-level segmentation were comparable with their fully supervised counterparts.

The contributions of this paper can be summarized as follows:
\begin{itemize}

  \item We propose a weakly supervised learning framework, CAMEL, for histopathology image segmentation using only image-level labels. CAMEL automatically enriches supervision information of the image by generating the instance-level labels from the image-level ones and achieves comparable performance with the fully supervised baselines in both instance-level classification and pixel-level segmentation.
  \item To construct a high-quality instance-level dataset for fully supervised learning, we introduce a cMIL approach which combines two complementary instance selection criteria (Max-Max and Max-Min) in the data preparation process to balance the data distribution in the constructed dataset.
  \item To fully utilize the original image-level supervision information, we propose the cascade data enhancement method and add image-level constraints to boost the performance of CAMEL further.
  \item To facilitate the research in histopathology field, our colorectal adenoma dataset will be made publicly available at https://github.com/ThoroughImages/CAMEL.

\end{itemize}

\section{Related Work}
\subsection{Weakly Supervision in Computer Vision}

In computer vision, people have proposed many weakly supervised algorithms~\cite{Ahn_2019_CVPR,Ahn_2018_CVPR,Durand2017WILDCAT,GeYY18,huang2018dsrg,wei2017object,wei2017stc} for object detection and semantic segmentation. However, in histopathology image analysis scenarios, the difference of morphological appearance between foreground (cancerous region) and background (non-cancerous region) is less significant~\cite{2019ScanNet} compared to what is usually observed in natural images. Moreover, the cancerous regions are disconnected, and their morphologies are usually various. Therefore, the methods based on adversarial erasing~\cite{wei2017object} or seed growing~\cite{Ahn_2018_CVPR} may not be suitable.

\subsection{Weakly Supervision in Histopathology Image}
\subsubsection{Instance-Level Classification}
MIL is widely applied in most weakly supervised methodologies~\cite{Jia2017Constrained,Xu2014Deep,Xu2014Weakly}. However, despite the great success of MIL, many solutions need pre-specified features~\cite{Viola2005Multiple,Xu2014Weakly}, which require data specific prior knowledge and limit the general applications. Instead of using pre-specified features, Xu et al.~\cite{Xu2014Deep} proposed to extract feature representations through a deep neural network automatically. However, the separation between feature engineering and MIL complicates the training process. In cMIL, the training procedure is end-to-end without deliberate feature extraction and feature learning, making the training process straightforward.

\subsubsection{Pixel-Level Segmentation}
Weakly supervised learning for histopathology image segmentation~\cite{Jia2017Constrained} has been proposed in recent years. The best performance was achieved by introducing artificial cancer area constraints. In CAMEL, the label enrichment step generates instance-level labels with more detailed supervision information and less labeling burden. In addition, compared to CDWS-MIL~\cite{Jia2017Constrained}, the classifier in CAMEL does not need pre-training and thus increases the flexibility in choosing the network architecture.

\section{Method}
\subsection{Label Enrichment}

Due to the lack of sufficient supervision information, simply using the image-level labels is insufficient to train the segmentation model. Therefore, before building the segmentation model, we perform a label enrichment procedure by generating instance-level labels from the original image-level labels (see Fig. \ref{f1}).

\subsubsection{Combined Multiple Instance Learning}

\begin{figure}[t]
\begin{center}
\includegraphics[width=0.45\textwidth]{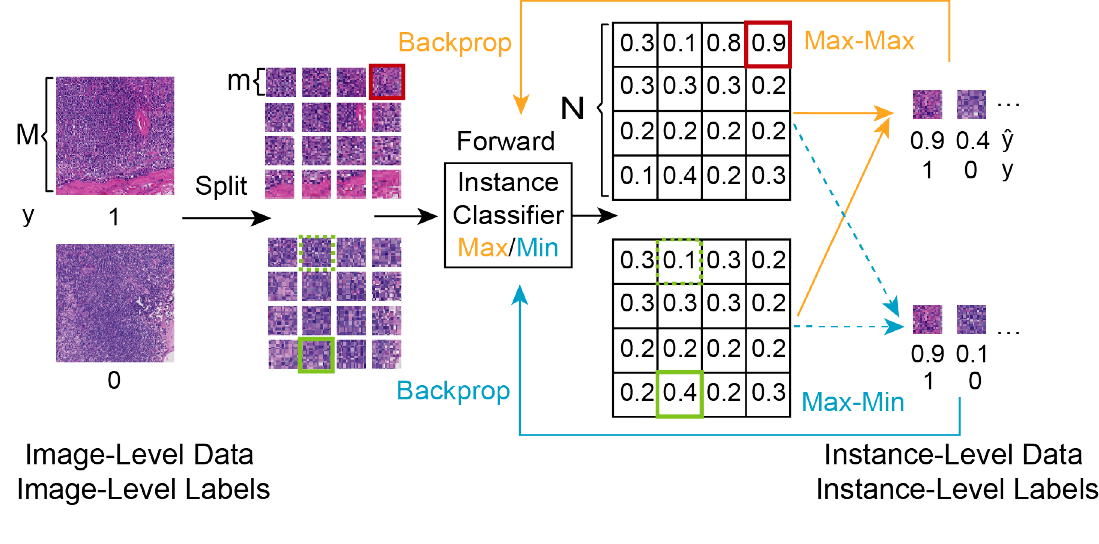}
\caption{Training procedure of cMIL. M and m represent the size of the image and the instance, respectively. $N$ is the \emph{scale factor} of cMIL where $N=\frac{M}{m}$, here we require $M$ to be divisible by $m$. We first split the image into $N \times N$ latticed instances with equal size. The selected instance can be considered as the representative of its corresponding image, therefore they own the same class label. We train two MIL models separately using two instance selection criteria (Max-Max and Max-Min).}
\label{f3}
\end{center}
\end{figure}

The effectiveness of CAMEL closely depends on the quality of our automatically enriched instance-level labels, which can also be regarded as a weakly supervised instance-level classification problem with only image-level labels. Here, we aim to transform this weakly supervised learning problem into a fully supervised instance-level classification one, and benefit from many existing well-developed fully supervised learning methods.

We introduce a new solution called \emph{combined Multiple Instance Learning} (cMIL). The image is split into $N \times N$ latticed instances with equal size. Here, we consider the instances from the same image as in the same \emph{bag}. In cMIL, two MIL-based classifiers with different instance selection criteria (Max-Max and Max-Min) are used to select instances to construct the instance-level dataset (Fig. \ref{f3}). The selected instance can be considered as the representative of its corresponding image, which determines the image class (similar to the attention mechanism~\cite{Xiao2014The}).

\begin{figure}[t]
\begin{center}
\includegraphics[width=0.45\textwidth]{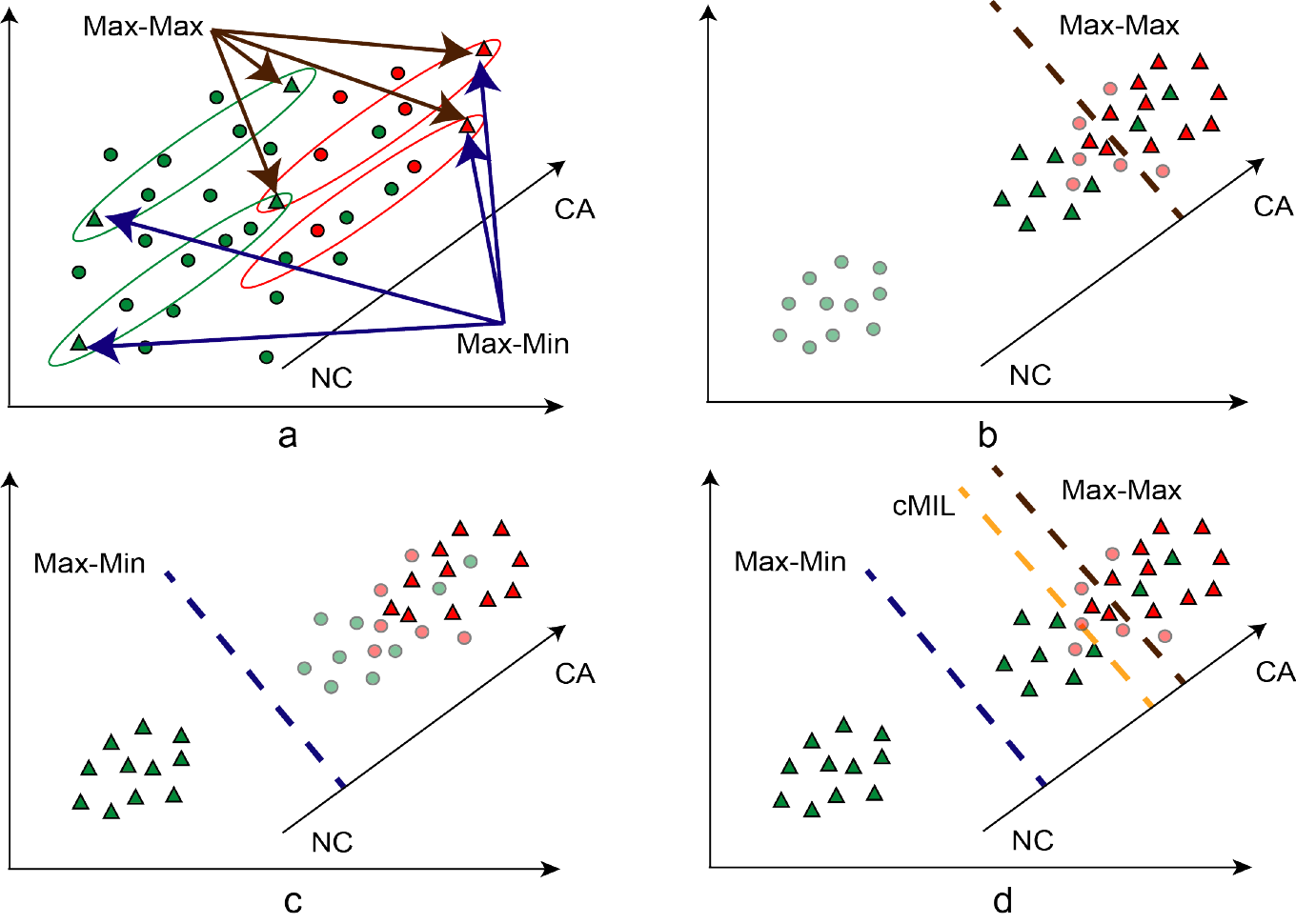}
\caption{Intuition behind two instance selection criteria named Max-Max and Max-Min. Red and green circles represent the CA and NC instances, respectively. We use triangles to represent the selected instances, and circles with light colors to represent the instances that are not selected. Each dotted line represents the decision boundary of the classifier, which is trained with the selected instances. Each ellipse represents an image (or a bag in MIL). cMIL, which combines Max-Max and Max-Min, achieves a better decision boundary.}
\label{f2}
\end{center}
\end{figure}

If the image has a cancerous (CA) region, we can reason that at least one instance is cancerous. On the other hand, if the label of the image is non-cancerous (NC), all the instances in it are non-cancerous. For both CA and NC images, Max-Max selects the instance with maximum CA response. As shown in Fig. \ref{f2}(a) and (b), during the training stage, in NC region, the Max-Max criterion will select the instance with maximum CA response, which has the highest similarity with CA, as the NC example. Therefore, the model trained with these data would give a decision boundary toward the CA direction, and this would lead to misclassification of CA instances with lower responses (as shown by light red circles). For example, CA instances with similar morphological appearances to NC may get misclassified. Max-Min acts as a countermeasure that selects the instances with the highest CA response for CA images and the instances with the lowest response for NC images. As shown in Fig. \ref{f2}(c), Max-Min tends to have an opposite effect compared to Max-Max. Therefore, in cMIL we combine these two criteria to reduce the distribution deviation problem and obtain a more balanced instance-level dataset to be used in fully supervised learning (see Fig. \ref{f2}(d)). It is worth noting that, for NC images, although each instance is NC, we only use the selected instances to avoid the data imbalance problem.

We choose ResNet-50~\cite{he2016deep} as the classifier. The two MIL-based classifiers are trained separately under the same configuration (Fig. \ref{f3}): in the forward pass, we use the Max-Max (or Max-Min for the other classifier) criterion to select one instance from each bag based on their predictions, and the prediction of the selected instance is regarded as the prediction of the image. In the backprop, we use the cross entropy loss between the image-level label and the prediction of the selected instance to update the classifier's parameters. The loss function for each classifier is defined as follows:

\begin{equation}
Loss = -\sum_{j}(y_j\log \hat{p_j}+(1-y_j)\log(1-\hat{p_j})),
\end{equation}
where $\hat{p_j}=S_{criterion}(\{f(b_i)\})$, $b_i$ is instances in image $j$, $f$ is the classifier, $S_{criterion}\in$ \{Max-Max, Max-Min\}. $S_{criterion}$ selects the target instance using the defined criterion, $y_j$ is the image-level label.

For Max-Max criterion:
\begin{equation}
S_{Max-Max}(\{f(b_i)\}) = \max \limits_{i}\{f(b_i)\}.
\end{equation}

For Max-Min criterion:
\begin{equation}
S_{Max-Min}(\{f(b_i)\}) = \left\{\begin{matrix}
\max \limits_{i}\{f(b_i)\} & if  \ y=1\\
\min \limits_{i}\{f(b_i)\} & if  \ y=0
\end{matrix}\right..
\end{equation}

After training, we again feed the same training data into the two trained classifiers and select the instances under the corresponding criterion, then the predictions are considered as their labels. We combine the instances selected by the two trained classifiers to construct the final fully supervised instance-level dataset. Noted that we discard those potentially confusing samples whose predicted labels are different from their corresponding image-level labels.

\subsubsection{Retrain and Relabel}

Once the instance-level dataset is prepared, we are able to train an instance classifier in a fully supervised manner. The classifier we use in this step has the same architecture as the classifier in cMIL (ResNet-50), we name this step as \emph{retrain}. Then, we split the original image into latticed instances and \emph{relabel} them using the trained instance-level classification model (Fig. \ref{f1}). For each image, we obtain $N^2$ high-quality instance labels from a single image-level label.

\subsection{Segmentation}

With enriched supervision information, the instance-level labels are directly assigned to the corresponding pixels, producing approximate pixel-level labels. Therefore, we can train segmentation models in a fully supervised way using well-developed architectures such as DeepLabv2~\cite{Chen2014Semantic,chen2018deeplab} and U-Net~\cite{Ronneberger2017U}. To prevent the model from learning the checkboard-like artifacts in the approximate labels, in the training process, we perform data augmentation by feeding smaller images that are randomly cropped from the original training set and their corresponding masks into the segmentation model.

\subsection{Further Improvement}


The granularity of the enriched labels is determined by the scale factor $N$; larger scale factor results in finer labels. However, as a tradeoff, larger scale factor would lead to severe image information loss. To tackle this issue, we propose cascade data enhancement to recover the potential loss and add image-level constraints to make better use of the supervision information.

\subsubsection{Cascade Data Enhancement}

\begin{figure}[t]
\begin{center}
\includegraphics[width=0.45\textwidth]{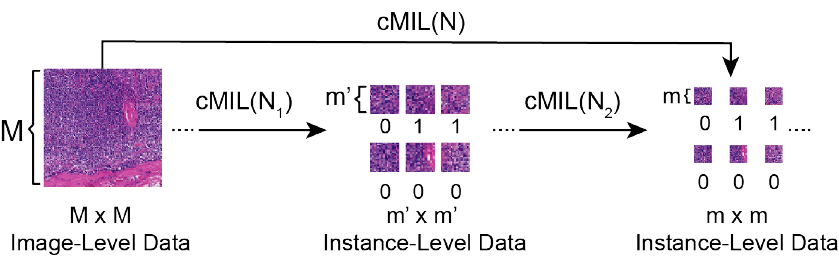}
\caption{Cascade data enhancement. Beside constructing the $m \times m$ dataset using cMIL($N$) directly, we can also first construct an intermediate ${m}'\times{m}'$ dataset using cMIL($N_1$), then construct the final m$\times$m dataset using cMIL($N_2$) in a cascade manner ($N=N_1 \times N_2$).}
\label{cascade}
\end{center}
\end{figure}

Each instance selection criterion only choose one instance from the image to construct the instance-level dataset, which only takes up a small portion of the image, resulting in losing a considerable amount of image information from the original image-level dataset. In order to recover this information loss and increase data diversity in the instance-level dataset, we further introduce the cascade data enhancement method to generate the instance-level dataset by two concurrent routes (Fig. \ref{cascade}). Here, we use cMIL($N$) to denote the cMIL with a scale factor of $N$. To derive labeled instances of a scale factor of $N$, we can either use cMIL($N$) or cMIL($N_1$) and cMIL($N_2$) back-to-back where $N=N_1 \times N_2$. The two sources of data are combined before fed into the segmentation model.

\subsubsection{Training with Image-Level Constraints}

\begin{figure*}[t]
\begin{center}
\includegraphics[width=0.75\textwidth]{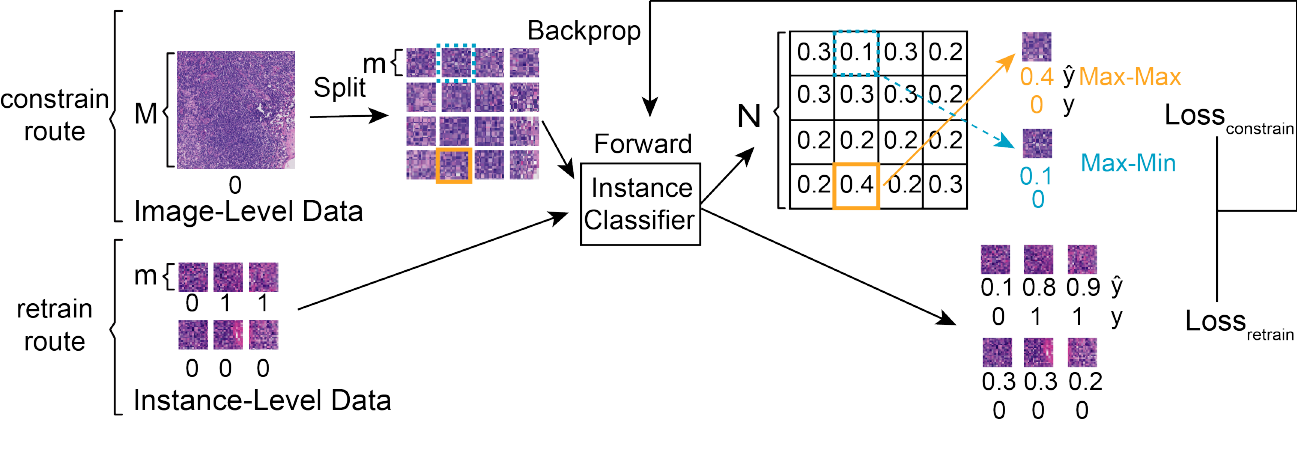}
\caption{Illustration of model training under image-level constraints. The supervision information from the original image-level data is taken into consideration in the retrain step.}
\label{fms}
\end{center}
\end{figure*}

In order to maximize the utility of the original image-level supervision information, in the retrain step, we can further add the original image-level data as one additional input source going through the classifier. As shown in Fig. \ref{fms}, the image-level constraint is imposed under Max-Max and Max-Min criteria to the instance level, the total loss is defined as the sum of the retrain loss and the constraint loss:
\begin{equation}
Loss = w_1 \cdot Loss_{constrain} + w_2 \cdot Loss_{retrain},
\end{equation}
where $w_1$ and $w_2$ are the weights of the two losses. We set $w_1 = w_2$ in our experiments.

\begin{equation}
Loss_{constrain} = -\sum_{S_{criterion}}(y\log \hat{p}+(1-y)\log(1-\hat{p})),
\end{equation}
where $\hat{p}=S_{criterion}(\{f(b_i)\})$, $b_i$ represents the selected instance, $f$ is the image-level constrain route, $S_{criterion}\in$ \{Max-Max, Max-Min\}, and $y$ is the image-level label.

\begin{equation}
Loss_{retrain} = -\sum_{j}(y_j\log \hat{y_j}+(1-y_j)\log(1-\hat{y_j})),
\end{equation}
where $\hat{y_j}=g(n_j)$, $n_j$ represents the input instance, $g$ is the retrain route, and $y_j$ is the instance-level label. Since two routes share the same network, we have $f \equiv g$.

\section{Experiments}
\subsection{Data Preparation}

We conducted our experiments on CAMELYON16~\cite{Camelyon2016,Ehteshami2017Diagnostic}, a public dataset with 400 hematoxylin-eosin (H\&E) stained whole-slide images (WSIs) of lymph node sections. In this research, same as CDWS-MIL~\cite{Jia2017Constrained}, we regard the 1,280$\times$1,280 patches at 20x magnification in the WSIs as image-level data. The training set of CAMELYON16 contains 240 WSIs (110 contain CA), which we split into 5,011 CA and 96,496 NC 1,280$\times$1,280 patches, and we over-sample the CA patches to match the number of NC ones. Besides, we have also constructed two other fully supervised training sets containing 320$\times$320 and 160$\times$160 instances to help build the fully supervised baselines. The test set includes 160 WSIs (49 contain CA), and we split and select all the 3,392 1,280$\times$1,280 CA patches, and then we randomly sample NC patches to match the number \footnote{We exclude Test\_114 because of the duplicate labeling~\cite{Yi2018Cancer}.}. The 1,280$\times$1,280 patches are further split into sizes of 320$\times$320 and 160$\times$160 to test the models with corresponding input sizes. The patches and the instances are labeled as CA if it contains any cancerous region. Otherwise, the label is NC.

\subsection{Implementation}
We applied rotation, mirroring, and scaling (between 1.0x and 1.2x) at random to augment the training data. All the models were implemented in TensorFlow \cite{abadi2016tensorflow} and trained on 4 NVIDIA GTX1080Ti GPUs. Both instance classifiers in cMIL and the retrain step were trained using Adam optimizer with a fixed learning rate of 0.0001. In cMIL, the batch size was set to 4 (one image-level patch on each GPU). In the retrain step, the batch size was set to 40 (ten instances on each GPU). During the segmentation stage, DeepLabv2 and U-Net were both trained using Adam optimizer with a fixed learning rate of 0.001 and the batch size of 24 (six images on each GPU). Due to the limitation of the GPU resources, we used 640$\times$640 images that are randomly cropped from the original 1280$\times$1280 training set and their corresponding masks to train the segmentation models.

\subsection{Performance of Label Enrichment}

\begin{table}
\begin{center}
\caption{Instance-level classification performance of label enrichment on CAMELYON16 test set.}
\label{t2}
\resizebox{0.45\textwidth}{!}{
\begin{tabular}{|l|c|c|c|}
\hline
320$\times$320 (\%)                         & Sensitivity  & Specificity & Accuracy \\
\hline
FSB320                                  & 90.0      & 97.4      & 94.5 \\
Max-Max                                 & 56.9      & 98.1      & 81.9 \\
Max-Min                                 & 82.0      & 82.6      & 82.3 \\
Retrain (cMIL)                          & 88.7      & 94.6      & {\bfseries 92.3} \\
Retrain (constrained)                   & 84.5      & 98.4      & {\bfseries 92.9} \\
\hline
\hline
160$\times$160 (\%)                         & Sensitivity  & Specificity & Accuracy \\
\hline
FSB160                                  & 89.0      & 95.0      & 92.8 \\
Max-Max                                 & 44.9      & 99.3      & 79.3 \\
Max-Min                                 & 87.7      & 86.5      & 86.9 \\
Retrain (cMIL)                          & 85.5      & 90.1      & {\bfseries 88.4} \\
Retrain (constrained)                   & 75.2      & 98.5      & 89.9 \\
Cascade                                 & 87.7      & 92.0      & 90.4 \\
Cascade (constrained)                   & 83.6      & 96.4      & {\bfseries 91.7} \\
\hline
\end{tabular}
}
\end{center}
\end{table}

\begin{figure*}
\begin{center}
\includegraphics[width=0.75\textwidth]{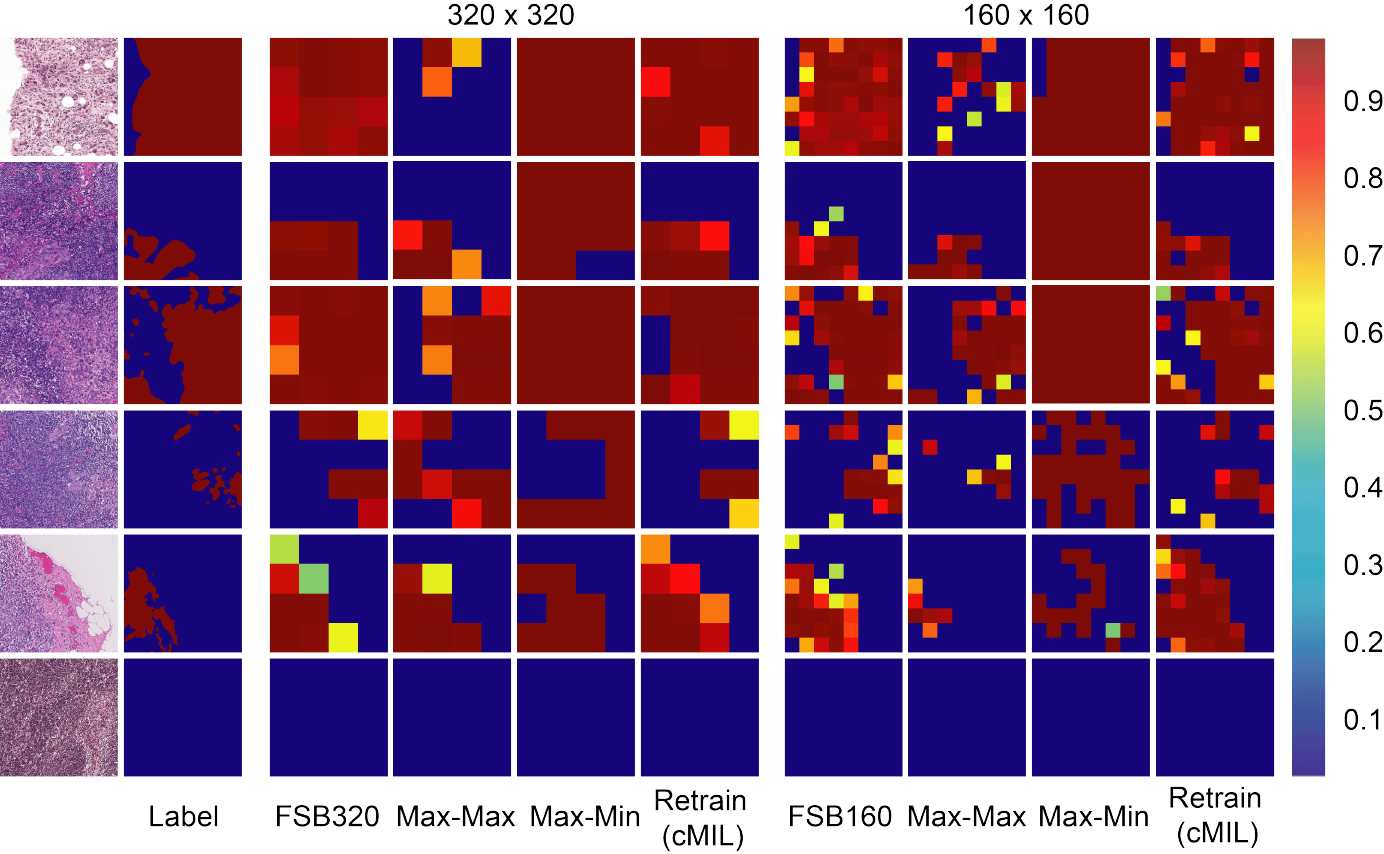}
\caption{Instance-level classification results on CAMELYON16 test set. Compare to the ground truth, the model trained on the data selected using Max-Max tends to predict less CA, and more CA using Max-Min. Retrain (cMIL) achieves a more reasonable trade-off and better performance.}
\label{f4}
\end{center}
\end{figure*}

As Table \ref{t2} and Fig. \ref{f4} show, in accordance with Fig. \ref{f2}, models trained on data selected using Max-Max tends to have relatively low sensitivity and high specificity. On the contrary, Max-Min tends to help achieve relatively high sensitivity and low specificity. With the data selected with the two criteria combined, the model can achieve a more reasonable trade-off and better performance. By using the cascade data enhancement method and adding the image-level constraints, we further improve the model's accuracy. To compare the performance between our model and the fully supervised baseline (FSB), we use the same classifier architecture (ResNet-50) for both models. On the 320$\times$320 and the 160$\times$160 test sets, the instance classification accuracy are comparable with the fully supervised baselines, which are only 1.6\% and 1.1\% lower, respectively.

The improvement from cascade data enhancement shows an effective way to recover from image information dilation in constructing the fully supervised instance-level dataset and suggests its potential for label enrichment on an even finer granularity. It also implicates the robustness of cMIL with different scale factors. The improvement from adding the image-level constraints shows the benefit of combining supervision information of image-level and instance-level.

We further verify the instance-level classification performance of our best models on the 320$\times$320 and 160$\times$160 training sets (Table \ref{t4}), where they achieve 95.5\% and 94.6\% accuracies, respectively. After label enrichment, CAMEL successfully enriches the supervision information from single image-level label to $N^2$ instance-level granularity for the images in the original image-level dataset with high quality.

\begin{table}
\begin{center}
\caption{Quality of automatically enriched instance-level labels for the original image-level dataset measured by the classification performance on CAMELYON16 training sets.}
\label{t4}
\resizebox{0.45\textwidth}{!}{
\begin{tabular}{|l|c|c|c|c|}
\hline
                    &$N^2$          & Sensitivity  & Specificity & Accuracy \\
\hline
160$\times$160      &64         & 89.9      & 94.7      & 94.6 \\
320$\times$320      &16         & 91.4      & 95.7      & 95.5 \\
\hline
\end{tabular}
}
\end{center}
\end{table}

\subsection{Performance of Segmentation}

\begin{table*}
\begin{center}
\caption{Pixel-level segmentation performance on CAMELYON16 test set.}
\label{t5}
\resizebox{0.75\textwidth}{!}{
\begin{tabular}{|l|c|c|c|c|c|}
\hline
DeepLabv2 (\%)                         & Sensitivity  & Specificity & Accuracy & F1-Score & IoU\\
\hline
Pixel-Level FSB                     &87.9    &99.1    &95.3    &92.6    &86.3 \\
Image-Level FSB                     &89.2    &88.7    &88.9    &84.4    &72.9 \\
CAMEL (160)                         &92.7    &95.7    &94.7    &92.1    &{\bfseries 85.4} \\
CAMEL (320)                         &94.7    &93.8    &94.1    &91.5    &{\bfseries 84.3} \\
\hline
\hline
U-Net (\%)                               & Sensitivity  & Specificity & Accuracy & F1-Score & IoU\\
\hline
Pixel-Level FSB                     &87.8    &98.2    &94.7    &91.8    &84.8 \\
Image-Level FSB                     &95.5    &82.1    &86.6    &82.8    &70.6 \\
CAMEL (160)                         &94.7    &94.1    &94.3    &91.8    &{\bfseries 84.8} \\
CAMEL (320)                         &94.7    &94.0    &94.2    &91.7    &{\bfseries 84.7} \\
\hline
\hline
Other Methods (\%)                      & Sensitivity  & Specificity & Accuracy & F1-Score & IoU\\
\hline
WILDCAT (w/ ResNet-50)              &69.6    &93.8    &85.7     &76.6    &62.0 \\
DWS-MIL (w/ ResNet-50)              &86.0    &93.4    &90.9     &86.4    &76.0 \\
CDWS-MIL (w/ ResNet-50)             &87.2    &93.8    &91.5     &87.4    &77.6 \\
\hline
\end{tabular}
}
\end{center}
\end{table*}

After label enrichment, the instance-level labels of the training set are assigned to the corresponding pixels to produce approximate pixel-level labels. At this point, we can train the segmentation model in a fully supervised manner. We test the performance of DeepLabv2 with ResNet-34 \cite{chen2018deeplab} and U-Net \cite{Ronneberger2017U}.

\begin{figure}
\begin{center}
\includegraphics[width=0.45\textwidth]{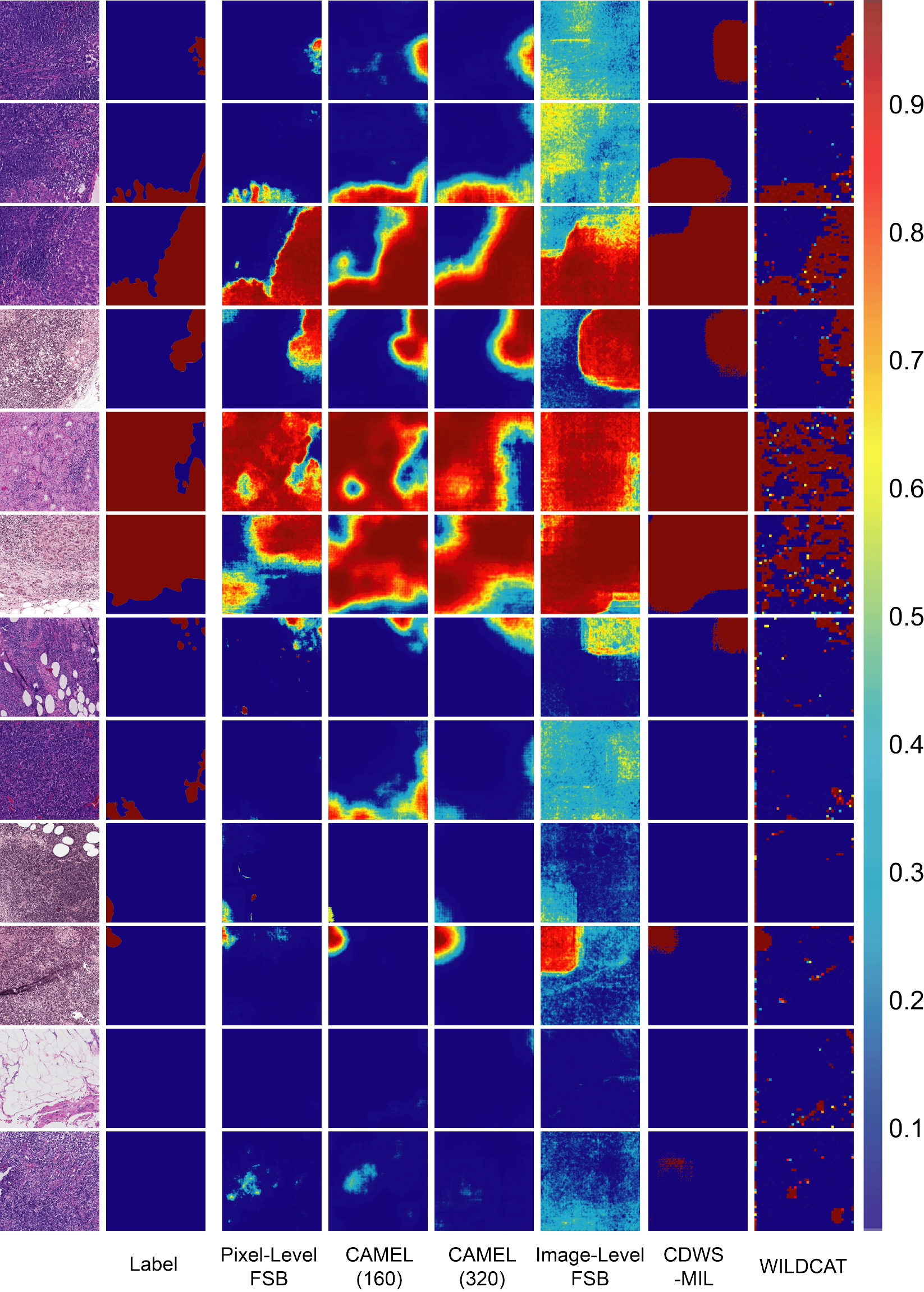}
\caption{Pixel-level segmentation results (DeepLabv2) of CAMEL and other methods on CAMELYON16 test set.}
\label{f6}
\end{center}
\end{figure}

As given in Table \ref{t5}, we use sensitivity, specificity, accuracy, and intersection over union (IoU) to measure the pixel-level segmentation performance. For comparison, the performance of the fully supervised baseline pixel-level FSB and the performance of the weakly supervised methods WILDCAT~\cite{Durand2017WILDCAT}, DWS-MIL, and CDWS-MIL~\cite{Jia2017Constrained} are also listed. WILDCAT is used for natural images in their paper~\cite{Durand2017WILDCAT}, and DWS-MIL and CDWS-MIL~\cite{Jia2017Constrained} are used for histopathology image. Here, we add another baseline model (image-level FSB) to show the importance of label enrichment for segmentation performance. The image-level FSB is trained with the data whose label is generated by directly assigning the image-level labels to the pixels, while the pixel-level FSB is obtained using the original pixel-level ground truth. CAMEL outperforms the image-level FSB, WILDCAT, DWS-MIL, and CDWS-MIL, and is even comparable with the pixel-level FSB.

With the help of the efficient use of supervision information, finer granularity brings with better segmentation performance. Moreover, in the label enrichment step, the instance pixels are labeled as CA if it contains any cancerous region. This may lead to the over-labeling issue. As shown in Fig. \ref{f6}, smaller instance size alleviates this issue by constructing finer pixel-level labels, demonstrating the effectiveness of finer labels and the potential of improvement for label enrichment on an even finer granularity.

\begin{figure*}
\begin{center}
\includegraphics[width=0.78\textwidth]{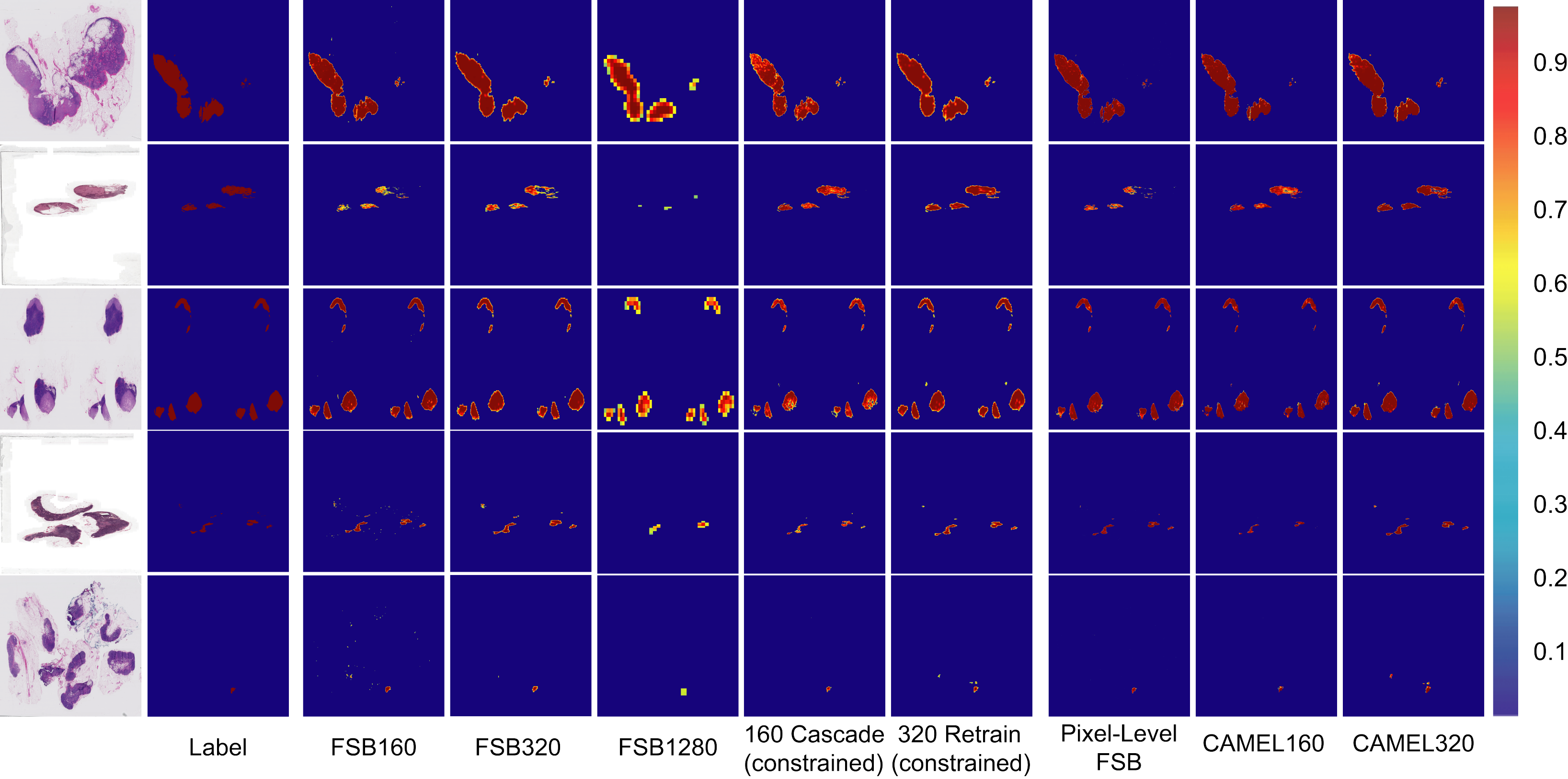}
\caption{Some examples of instance-level classification and pixel-level segmentation (DeepLabv2) results on CAMELYON16 WSIs.}
\label{f5}
\end{center}
\end{figure*}

We further evaluate our models on the WSIs of CAMELYON16 test set. Fig. \ref{f5} shows some examples.

\subsection{Generality of CAMEL}

\begin{table}
\begin{center}
\caption{Model performance on colorectal adenoma dataset.}
\label{ts1}
\resizebox{0.48\textwidth}{!}{
\begin{tabular}{|l|c|c|c|}
\hline
Instance-level classification (\%)      & Recall  & Precision & Accuracy \\
\hline
FSB320                                  & 81.1      & 90.0      & 87.1 \\
Retrain (cMIL)                          & 84.9      & 81.0      & 83.8 \\
FSB160                                  & 80.7      & 87.6      & 87.0 \\
Retrain (cMIL)                          & 80.9      & 85.1      & 86.0 \\
\hline
\hline
Pixel-level segmentation (\%)           & Recall    & Precision & F1-Score \\
\hline
Pixel-Level FSB                         &86.1       &89.0       &87.5 \\
CAMEL (160)                             &89.7       &85.0       &87.3 \\
CAMEL (320)                             &95.4       &78.5       &86.1  \\
\hline
\end{tabular}
}
\end{center}
\end{table}

\begin{figure}
\begin{center}
\includegraphics[width=0.42\textwidth]{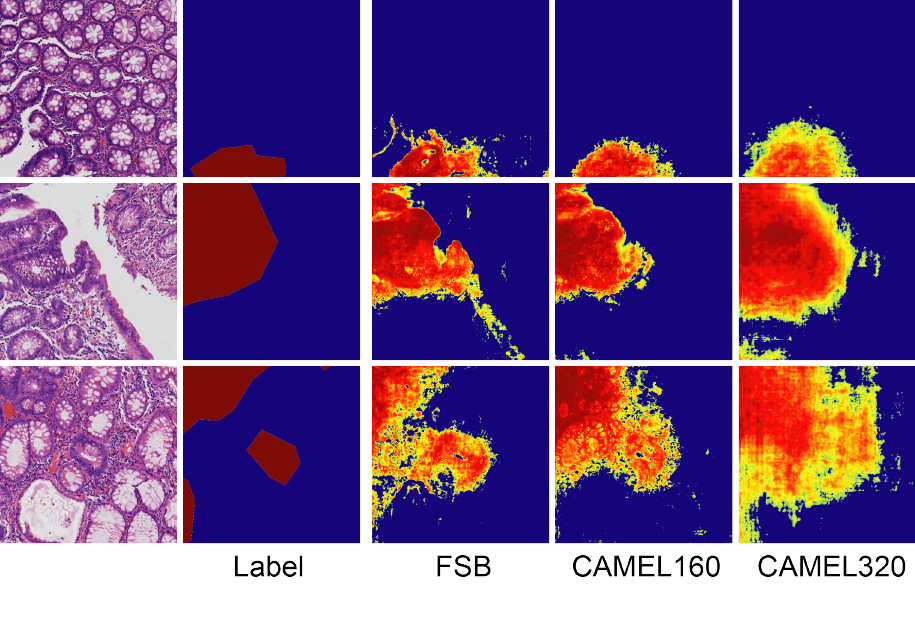}
\caption{Pixel-level segmentation results (DeepLabv2) of CAMEL on colorectal adenoma dataset.}
\label{fs1}
\end{center}
\end{figure}

To evaluate the generality of CAMEL, we test CAMEL on a colorectal adenoma dataset which contains 177 WSIs (156 contain adenoma) gathered and labeled by pathologists from the Department of Pathology, The Chinese PLA General Hospital. As Table \ref{ts1} and Fig. \ref{fs1} show, CAMEL consistently achieves comparable performance against the fully supervised baselines.

\section{Conclusion}

Computer-assisted diagnosis for histopathology image can improve the accuracy and relieve the burden for pathologists at the same time. In this research, we present a weakly supervised learning framework, CAMEL, for histopathology image segmentation using only image-level labels. CAMEL automatically enriches supervision information from image-level to instance-level with high quality and achieves comparable segmentation results with its fully supervised counterparts. More importantly, the automatic labeling methodology may generalize to other weakly supervised learning studies for histopathology image analysis.

In CAMEL, the obtained instance-level labels are directly assigned to the corresponding pixels and used as masks in the segmentation task, which may result in the over-labeling issue. We will tackle this challenge using mask boundary refinement~\cite{Ahn_2019_CVPR,Ahn_2018_CVPR} in future work.

\paragraph{Acknowledgement.}
The authors would like to thank Xiang Gao, Lang Wang, Cunguang Wang, Lichao Pan, Fangjun Ding at Thorough Images for data processing and helpful discussions. This research is supported by National Natural Science Foundation of China (NSFC) (No. 61532001), Tsinghua Initiative Research Program (No. 20151080475), Shanghai Municipal Science and Technology Major Project (No. 2018SHZDZX01) and ZJLab.

{\small
\bibliographystyle{ieee_fullname}
\bibliography{camel}
}

\end{document}